\documentclass[12pt]{article}

\usepackage{graphicx}
\usepackage{amsmath,amssymb,amsfonts,multirow,verbatim,tabularx,slashed}
\usepackage{times}
\usepackage{youngtab}
\usepackage{mathbbol}

\newcommand{\PB}[2]{\left\{#1,\,#2\right\}_{\mathrm{PB}}}
\newcommand{\COM}[2]{\left[#1,\,#2\right]}
\newcommand{\SCOM}[2]{\left[#1,\,#2\right\}}

\newcommand{\tr}{{\sf tr}}
\newcommand{\g}{{\sf g}}
\renewcommand{\div}{{\sf div}}
\newcommand{\grad}{{\sf grad}}
\renewcommand{\outer}{\iota_\xivec^*{}}
\newcommand{\inner}{\iota_\xivec^{\phantom{*}}{}}
\newcommand{\ord}{{\sf ord}}
\newcommand{\N}{{\sf N}}
\newcommand{\cas}{{\sf c}}
\newcommand{\curl}{{\sf curl}}
\newcommand{\abox}{\blacksquare}
\newcommand{\boxop}{\square}
\newcommand{\xivec}{{\boldsymbol{\xi}}}
\newcommand{\be}{\begin{equation}}
\newcommand{\ee}{\end{equation}}

\newcommand{\liefont}[1]{\mathfrak{#1}}
\renewcommand{\sp}{\liefont{sp}}
\newcommand{\osp}{\liefont{osp}}
\newcommand{\dbddx}[1]{dx^*_{#1}}
\newcommand{\dbddxs}[1]{\dbddx{#1}}

\newcommand{\covartime}[1]{\frac{\nabla #1}{dt}}

\newcommand{\covarvar}{\mathcal{D}}

\renewcommand{\Gamma}{\varGamma}

\def\clap#1{\hbox to 0pt{\hss#1\hss}}
\def\mathllap{\mathpalette\mathllapinternal}

\def\mathllapinternal#1#2{%
\llap{$\mathsurround=0pt#1{#2}$}}

\begin{document}

\title{\sc Conformal Orthosymplectic Quantum Mechanics}

\author{\begin{tabular}{ccc}
Joshua Burkart && Andrew Waldron\\
Department of Physics && Department of Mathematics\\
University of California&&University of California\\
Berkeley CA 94720 && Davis CA 95616\\
\tt{\small burkart@berkely.edu}&&\tt{\small wally@math.ucdavis.edu}
\end{tabular}
}
%

\maketitle

\begin{abstract}
\noindent
We present the most general curvature obstruction to the deformed parabolic 
orthosymplectic symmetry subalgebra of the supersymmetric quantum mechanical models recently developed
to describe Lichnerowicz wave operators acting on arbitrary  tensors and spinors.
For geometries possessing a
hyper\-sur\-face-or\-thog\-onal homothetic conformal Killing vector 
we show that the parabolic subalgebra is enhanced to a (curvature-obstructed) ortho\-symplectic
algebra. The new symmetries correspond to time-dependent conformal symmetries
of the underlying particle model. 
We also comment on generalizations germane
to three dimensions and new Chern--Simons-like particle models.

\end{abstract}

\newpage

\tableofcontents

\section{Introduction}

Quantum mechanical models have been developed recently whose wavefunctions correspond to sections of general tensor and spinor bundles. Noeth\-er charges in these models describe operators such as the Laplacian, gradient, divergence, trace, exterior derivative, and Dirac operator \cite{Hallowell:2007qk,Hallowell:2007zb}. The symmetry algebra of these Noether charges  yields algebras for differential-geometric operators. Classically these models correspond to particles with internal degrees of freedom, whose evolution generalizes parallel transport and geodesic motion. In~\cite{Hallowell:2007qk,Hallowell:2007zb} the main focus was on locally symmetric (pseudo)Rie\-mann\-ian manifolds in order that a Lich\-ner\-o\-wicz-type Laplacian or wave operator \cite{Lichnerowicz:1964zz} would be a central charge, which could therefore be taken as the Hamiltonian of the underlying quantum mechanical model, so that in turn all other operators would correspond to symmetries of the theory. As a result, the authors of \cite{Hallowell:2007qk,Hallowell:2007zb} found symmetries that formed a parabolic subalgebra of the super Lie algebra $\osp(Q|2p+2)$. The main aim of this work is to extend that subalgebra to the full orthosymplectic algebra $\osp(Q|2p+2)$. We will achieve this goal by considering curved backgrounds with conformal symmetry.

The models we consider generalize many well-known theories and have a wide range of applications.
For all curved backgrounds, they enjoy an $\osp(Q|2p)$ internal symmetry. The series of models $\osp(Q|0)$ correspond to $\liefont{o}(Q)$ spinning particle models \cite{Gershun:1979fb,Howe:1988ft}. The lowest cases $Q=1,2$ are
the ${\cal N}=1,2$ supersymmetric quantum mechanical models originally employed by Alvarez-Gaum\'e and Witten in their study of gravitational anomalies and Pontryagin classes~\cite{AlvarezGaume:1983ig}, and by Witten in an application to Morse theory~\cite{Witten:1982im}.
(Indeed, the Hodge-Lefschetz symmetry algebra of ${\cal N}=4$ supersymmetric quantum mechanics in K\"ahler backgrounds~\cite{Witten:1982df,FigueroaO'Farrill:1997ks} plays an analogous
role to the $\osp(Q|2p)$ algebras studied here.)

The dynamical, as opposed to internal, $R$-symmetries of our models are sensitive
to details of the background geometry. Geometrically, they correspond to 
linear differential operators such as the gradient, divergence, exterior
derivative, codifferential, Dirac operator, and generalizations thereof. These operators commute with the
Lichnerowicz wave operator  when the background is locally symmetric.
(In fact, in the case of the $\osp(1|0)$ and $\osp(2|0)$ models, the Lichnerowicz wave operator
is central in {\it any} curved background.)
In the most general curved backgrounds, however, there is a curvature obstruction to
dynamical symmetries. An important new result of this paper is an explicit computation
of this obstruction appearing as the result of commutators between the
Lichnerowicz wave operator and the linear, dynamical symmetry operators.

Armed with the generalization of the results of~\cite{Hallowell:2007qk} to arbitrary backgrounds,
we can investigate non-symmetric spaces. Because our philosophy is to develop models
which maximize symmetries, we consider spaces which share many features of flat space.
Already in~\cite{Hallowell:2007qk} it was observed that together, the dynamical and internal symmetries of the symmetric space models formed a parabolic Lie subalgebra of a larger $\osp(Q|2p+2)$ superalgebra
if one introduced a new operator which measured the engineering dimension of the existing symmetry charges. This strongly suggests the study of quantum mechanical models with conformal
symmetry. The first such model was developed some time ago in~\cite{de Alfaro:1976je}.
In particular, it was shown that the dilation operator corresponded to a time-\emph{dependent}
symmetry of the underlying particle action. There is an extensive literature concerning
supersymmetric generalizations of conformal quantum mechanics; in particular,
for the case that the dynamical symmetries are supersymmetries, the
authoritative study of~\cite{Papadopoulos:2000ka} is invaluable (in fact the current paper 
could be viewed as the synthesis of~\cite{Hallowell:2007qk} and~\cite{Papadopoulos:2000ka}).

In flat backgrounds the dilation operator corresponds to the Euler operator, which generates radial dilations. This property can be mimicked in more general backgrounds by requiring the existence
of a hyper\-surface-orthogonal homothetic conformal Killing vector. Flat space is the only symmetric space satisfying this requirement. Therefore, in flat backgrounds, the $\osp(Q|2p)$ models
in fact enjoy a larger, time-dependent $\osp(Q|2p+2)$ symmetry algebra that contains
an $\liefont{sl}(2,{\mathbb R})$ subalgebra, which is precisely the conformal algebra
of the one-dimensional particle worldline. However, in general backgrounds obeying
the hypersurface orthogonal homothetic conformal Killing vector condition the $\osp(Q|2p+2)$
algebra remains largely unscathed save for exactly the curvature obstruction discussed above.

The paper is organized as follows. In Section~\ref{sectgeometry} we give the relevant mathematical details for manifolds possessing
hypersurface orthogonal,  homothetic, conformal Killing vectors. In Section~\ref{sectsp4} we consider the theory of symmetric tensors
on such backgrounds and reformulate it as a quantum mechanical system with an $\sp(4)$ symmetry. Then in Section~\ref{sectosp}
we extend this analysis to tensors of arbitrary type and show how to describe them by a supersymmetric quantum mechanical
system with $\osp(Q|2p)$ symmetry. In Section~\ref{three} we specialize to three dimensions where we can extend our algebras by the 
symmetrized curl operation discovered in~\cite{Damour:1987vm}.

\section{Geometry}
\label{sectgeometry}

Our results apply to curved, torsion-free, $d$-dimensional, (pseudo)Rie\-mann\-ian manifolds.\footnote{We label flat indices with Latin letters $m$, $n$, etc., and curved indices with Greek letters $\mu$, $\nu$, etc. Greek letters $\alpha$, $\beta$, $\gamma$,  $\delta$,... are reserved for  the orthosymplectic superindices. The covariant derivative is defined by example as
$\nabla_{\mu}v^{\nu n} = \partial_\mu v^{\nu n}+\Gamma^\nu_{\mu\rho}v^{\rho n}+\omega_\mu{}^n{}_r v^{\nu r}.
$
Our Riemann tensor conventions are summarized by:
\begin{displaymath}\begin{split}
R_{\mu\nu\rho}{}^{\sigma} v_{\sigma}& = \big[\nabla_{\mu}, \nabla_{\nu}\big]v_{\rho}
 = 2\left({\partial\vphantom{\Gamma}}_{[\nu}^{}\Gamma^{\sigma}_{\mu]\rho} + \Gamma^{\lambda}_{\rho[\mu}\Gamma^{\sigma}_{\nu]\lambda}\right)v_{\sigma}=e^r{}_\rho R_{\mu\nu r}{}^s v_s\\
&= 2\left(e^m{}_\mu e^n{}_\nu e^r{}_\rho\right)\left(\partial\vphantom{\omega}_{[m}\omega_{n]r}{}^s + \omega_{[mn]}{}^t\omega_{tr}{}^s + \omega_{[mr}{}^t\omega_{n]t}{}^s\vphantom{\Gamma^{\lambda}_{\rho[\mu}}\right)v_s.
\end{split}\end{displaymath}}
We will often specialize to spaces which possess a hypersurface-orthogonal homothetic conformal Killing vector, i.e., a vector field $\xivec=\xi^\mu\partial_\mu$ satisfying
\begin{equation}\label{eqhomothety}
g_{\mu\nu}=\nabla_{\mu}\xi_{\nu}.
\end{equation}
Henceforth we refer to the condition~\eqref{eqhomothety} as \emph{hyperhomothety}. From $\xivec$, we can form the homothetic potential
\begin{equation}
  \phi=\frac{\xi^{\mu}\xi_{\mu}}{2},
\end{equation}
which satisfies
\begin{equation}
  \nabla_\mu\phi=\xi_\mu,
\end{equation}
and consequently
\begin{equation}
  g_{\mu\nu}=\nabla_\mu\partial_\nu \phi.
\end{equation}

Under the hyperhomothety condition, it can be shown that the manifold admits coordinates $(r,x^i)$ such that the metric is explicitly a cone over some base manifold~\cite{Gibbons:1998xa} with metric $h_{ij}$, where in particular the homothetic potential is simply
\begin{equation}
  \phi=\frac{r^2}{2},
\end{equation}
and
\begin{equation}
ds^2=g_{\mu\nu}dx^{\mu}dx^{\nu}=dr^2+r^2 h_{ij}dx^{i}dx^{j}\, .
\end{equation}
Note that hyperhomothety~\eqref{eqhomothety} immediately implies that the
contraction of $\xivec$ on the Riemann tensor vanishes:
\begin{equation}
R_{\mu\nu\rho}{}^{\sigma}\xi_{\sigma}=\big[\nabla_{\mu},\nabla_{\nu}\big]\xi_{\rho}=0.
\end{equation}
The most elementary example of a hyperhomothetic space is flat space
\begin{equation}
ds^2=\eta_{\mu\nu}dx^\mu dx^\nu\quad\Rightarrow\quad\xivec=x^\mu\frac{\partial}{\partial x^\mu},
\end{equation}
where $\xivec$ is simply the Euler vector field---the dilation generator. In this sense, 
hyperhomothetic spaces are  very close to flat space and in many computations the components
$\xi^\mu$ of $\xivec$ mimic the flat space coordinates $x^\mu$.

Locally symmetric spaces,
\begin{equation}\label{eqsymmspace}
\nabla_{\lambda}R_{\mu\nu\rho\sigma}=0,
\end{equation}
enjoy isometries and are, in that sense, also similar to flat space.
This condition was explored in~\cite{Hallowell:2007qk} to obtain particle models with symmetries subject to a maximal parabolic subalgebra of the superalgebra $\osp$.

These two conditions---hyperhomothety in Eq.~\eqref{eqhomothety}, and locally symmetric space from Eq.~\eqref{eqsymmspace}---intersect only in flat space, and are otherwise \emph{mutually exclusive}. To see this,  apply Eq.~\eqref{eqhomothety} to obtain the identity
\begin{equation}
\xi^{\kappa}\nabla_{\kappa} R_{\mu\nu\rho\sigma}= -2 R_{\mu\nu\rho\sigma}\, .
\end{equation}
Clearly this contradicts the symmetric space condition~\eqref{eqsymmspace} unless the 
Riemann tensor vanishes.

In the following sections we will use the geometric conditions just presented to explore algebras
of differential geometry operators, which can in turn be reformulated as the quantization of orthosymplectic particle models. Our first example is the theory of symmetric tensors.

\section{$\sp(4)$ Conformal Quantum Mechanics}
\label{sectsp4}

Here we  present a set of geometric operators on totally symmetric tensors which form a representation of the $\sp(4)$ Lie algebra. We then show that these operators can  be reinterpreted as quantized Noether charges of a particle with intrinsic structure. These charges
correspond to rigid, continuous symmetries obeying the same~$\sp(4)$ algebra with an
$\sp(2)$ subalgebra playing the role of the worldline conformal group. This necessitates
both time-independent and dependent symmetries.

\subsection{Symmetric Tensors}\label{sectsp4opsymmtens}
Totally symmetric tensors can be represented as analytic functions by completely contracting all indices with commuting coordinate differentials~$dx^{\mu}$. Given a $\mbox{rank-$n$}$ symmetric tensor field 
$\phi_{\nu_1\nu_2\ldots\nu_n}$, we introduce
\begin{equation}
\Phi(x^{\mu},dx^{\mu})=\phi_{\nu_1\nu_2\ldots\nu_n}\!(x)\, dx^{\nu_1}dx^{\nu_2}\cdots dx^{\nu_n}.
\label{symtensor}
\end{equation}
Hence we can interpret the original tensor $\phi$  as a function $\Phi$ of the coordinates~$x^\mu$ and an \emph{analytic} function of commuting coordinate differentials $dx^\mu$. Note that it is now possible to add symmetric tensors of different ranks. In this indexless notation, one can then define various important operators\footnote{The algebra we present here was first discovered by Lichnerowicz~\cite{Lichnerowicz:1964zz} and then formalized in~\cite{Damour:1987vm}. It was subsequently employed in studies of
higher spin theories in~\cite{Labastida:1987kw,Vasiliev:1988xc,Hallowell:2005np}.}. To this end, in addition to coordinate differentials it is useful to introduce dual objects~$\dbddxs{\mu}$, which mutually commute but obey the Heisenberg algebra
\begin{equation}
\COM{\dbddx{\mu}}{dx^{\nu}}=\delta^{\nu}_{\mu}.\label{heis}
\end{equation}
Since we now consider coordinate differentials like coordinates, we can represent these dual 
differentials acting on symmetric tensors by 
\begin{displaymath}
\dbddx{\mu}=\frac{\partial}{\partial (dx^{\mu})}\, .
\end{displaymath}

With these ingredients we can build an operator which is equivalent to the covariant derivative when it acts on an symmetric tensor in the indexless form~\eqref{symtensor}, which we denote $D_\mu$ (distinct from the ordinary covariant derivative $\nabla_{\mu}$, which acts on tensors with indices):
\begin{equation}\label{eqsp4D}
D_{\mu}=\partial_{\mu}-\Gamma^{\sigma}_{\mu\nu}dx^{\nu}\dbddx{\sigma},
\end{equation}
where $\partial_\mu$ denotes the partial derivative $\partial/\partial x^{\mu}$. 
I.e., 
\begin{displaymath}
D_\mu\Phi=(\nabla_\mu \phi_{\nu_1\nu_2\ldots \nu_n}) \, dx^{\nu_1}dx^{\nu_2}\cdots dx^{\nu_n}\, .
\end{displaymath}
We can also build the Lor\-entz/rot\-ation generators
\begin{displaymath}
  M^{\mu\nu}=2g^{\rho[\nu}dx^{\mu]}\dbddx{\rho}\, ,\qquad [M^{\mu\nu},M_{\rho\sigma}]=4\, M
  ^{[\mu}\!\!\! {\phantom\delta}^{\phantom{\nu]}}_{[\sigma}\delta^{\nu]}_{\rho\hspace{.3mm} ]}\, ,
\end{displaymath}
where $[\cdot\,\cdot]$ denotes antisymmetrization with unit weight. Note that although $D_\mu$ and $M^{\mu\nu}$  act on tensors contracted with coordinate differentials, their outputs have \emph{open indices}. For this reason, they will not appear alone in the algebra we will discuss, but only in larger composite operators.

\begin{figure}[tbp]
\centering
\includegraphics{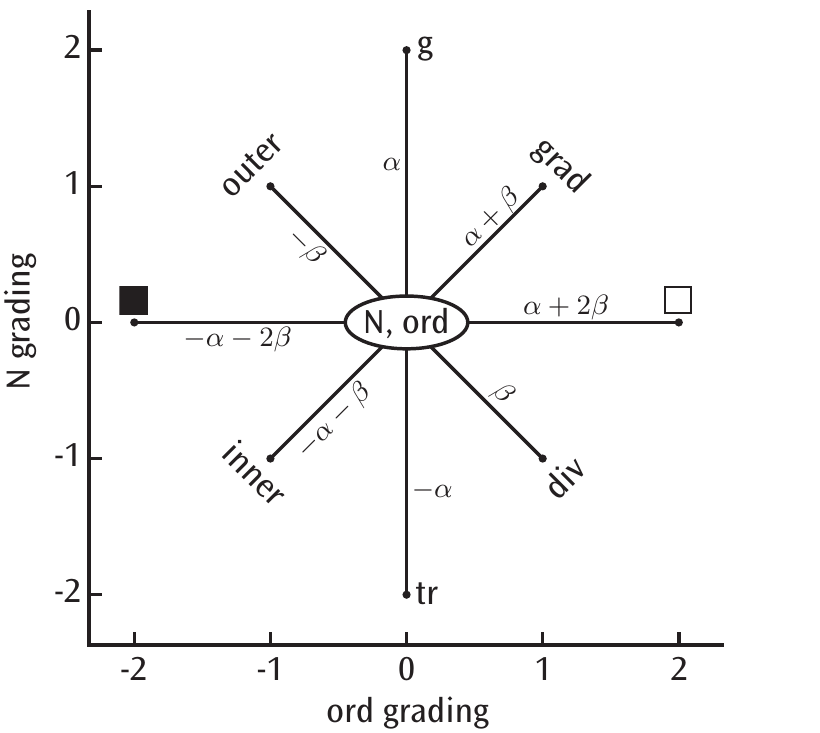}
\caption{Root lattice for the ten bilinear symmetric-tensor operators as a representation of $\sp(4)$, graded by the Cartan generators $\N$ and $\ord$. Their explicit commutation algebra is given in Appendix~\ref{appcomms}.} \label{figsp4rootlatt}
\end{figure}

\begin{table}[tbp]
\renewcommand{\arraystretch}{1.5}
\centering\begin{tabular}{|c|c|l|l|}
\hline \multicolumn{1}{|c|}{\it Name} & \it Root & \multicolumn{1}{c|}{\it Operator} & \multicolumn{1}{c|}{\it Interpretation}\\\hline
$\N$ & Cartan & $dx^{\mu}\dbddx{\mu}$ & Counts tensor rank\\\hline
$\g$ & $(0,2)$ & $g_{\mu\nu}dx^{\mu}dx^{\nu}$ & Metric outer product\\\hline
$\tr$ & $(0,-2)$ & $g^{\mu\nu}\dbddx{\mu}\dbddx{\nu}$ & Trace\\\hline
$\grad$ & $(1,1)$ & $dx^{\mu}D_{\mu}$ & Gradient\\\hline
$\div$ & $(1,-1)$ & $g^{\mu\nu}\dbddx{\mu}D_{\nu}$ & Divergence\\\hline
\multirow{2}{*}{$\boxop$} & \multirow{2}{*}{$(2,0)$} & \multirow{2}{*}{$\begin{aligned}&g^{\mu\nu}\left(D_{\mu}D_{\nu} - \Gamma_{\mu\nu}^{\sigma}D_{\sigma}\right)\\&+ R_{\mu}{}^{\nu}{}_{\rho}{}^{\sigma}dx^{\mu}\dbddxs{\nu}dx^{\rho}\dbddxs{\sigma}\end{aligned}$} &\multirow{2}{3.3cm}{Lichnerowicz wave operator} \\
&&&\\\hline
$\ord$ & Cartan & $-\xi^{\mu}D_{\mu}$ & Counts derivatives \\\hline
$\inner$ & $(-1,-1)$ & $\xi^{\mu}\dbddx{\mu}$ & Vector field inner product\\\hline
$\outer$ & $(-1,1)$ & $g_{\mu\nu}\xi^{\mu}dx^{\nu}$ & Vector field outer product\\\hline
$\abox$ & $(-2,0)$ & $g_{\mu\nu}\xi^{\mu}\xi^{\nu}$ & Scalar field multiplication\\\hline
\end{tabular}\
\renewcommand{\arraystretch}{1}
\caption{List of geometric symmetric-tensor operators, complete with root vectors for their corresponding $\sp(4)$ root lattice displayed in Figure~\ref{figsp4rootlatt}.}\label{tablesp4opsymmtens}
\end{table}

Now we will introduce a set of operators which map symmetric tensors to symmetric tensors (without producing extra indices). First, using just the operators $dx$ and $\dbddx{}$, we can construct three bilinears:
\begin{equation}\label{eqsp4ngtr}
\N=dx^\mu\dbddx{\mu},\quad \g=g_{\mu\nu}dx^\mu dx^\nu,\quad \tr=g^{\mu\nu}\dbddx{\mu}\dbddx{\nu}.
\end{equation}
Geometrically, these operators perform the following tasks: $\N$ determines tensor rank; $\g$ is the symmetrized outer product with the metric tensor; and $\tr$ is the trace/contraction with the metric tensor.

The above operators are ``non-dynamical'', whereas
employing the covariant operator $D_\mu$, we can form another three dynamical bilinears:
\begin{gather}
\grad=dx^\mu D_\mu,\quad \div=g^{\mu\nu}\dbddx{\mu}D_{\vphantom{\mu}\nu}^{\vphantom{*}},\notag\\\label{eqsp4diffops}
\begin{split}
\boxop&={\rm \Delta} + \frac{1}{4}R_{\mu\nu\rho\sigma}M^{\mu\nu}M^{\rho\sigma}.\\
\end{split}
\end{gather}
We are forced to add a ``quantum ordering'' term\footnote{For a study of operator orderings in supersymmetric quantum mechanics see~\cite{deAlfaro:1987mp}.} to form the Laplacian from the operator $D_\mu$
\begin{displaymath}
  {\rm \Delta}=g^{\mu\nu}\left(D_{\mu}D_{\nu} - \Gamma_{\mu\nu}^{\sigma}D_{\sigma}\right).
\end{displaymath}
We have added a curvature term to ${\rm \Delta}$ to make the operator $\boxop$ for symmetry reasons soon to become apparent. These operators can be interpreted as follows: $\grad$ is the symmetrized gradient; $\div$ is the symmetrized divergence; and $\boxop$ is the Lich\-nerowicz wave operator~\cite{Lichnerowicz:1964zz}. The operators from Eqs.~\eqref{eqsp4ngtr} and~\eqref{eqsp4diffops} are discussed more thoroughly in~\cite{Hallowell:2007qk}.

Lastly, using the  hyperhomothetic Killing vector $\xivec$, we can form four additional bilinears:
\begin{equation}\label{eqsp4conformalops}
  \ord=-\xi^\mu D_\mu,\quad\inner=\xi^\mu\dbddx{\mu},\quad \outer=\xi_\mu dx^\mu,\quad \abox=\xi^\mu\xi_\mu.
\end{equation}
These four operators can, of course, be defined whether or not the hyperhomothety 
condition holds; if it does not, $\xivec$ can be an arbitrary vector field, but if $\xivec$ obeys~\eqref{eqhomothety}  they have additional special properties. Indeed, these operators can be interpreted as follows: $\ord$ is the covariant derivative $\nabla_{\!\xivec}$ along $\xivec$---subject to hyperhomothety it counts derivatives. The operators $\inner$ and $\outer$ are, respectively,  the symmetrized inner and outer products with the vector field~$\xivec$.
Lastly, $\abox$ is multiplication by the scalar field/homothetic potential~$\xivec^2$.

\begin{figure}[htbp]
\renewcommand{\arraystretch}{1.5}
\newcolumntype{S}{>{\centering\arraybackslash} m{113pt} }
\newcolumntype{T}{>{\arraybackslash} m{1.2in} }
\centering\begin{tabular}{|r|S|T|T|}
\cline{2-4} \multicolumn{1}{c|}{} & \multicolumn{1}{c|}{\it Root lattice} & \multicolumn{1}{c|}{\it Conditions} & \multicolumn{1}{c|}{\it Algebra}\\\hline
i.&\includegraphics{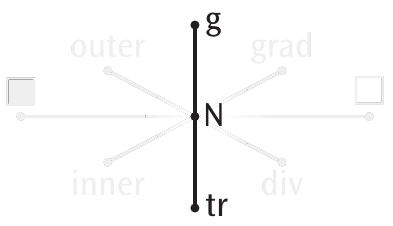} & \multirow{1}{*}{Any background} & \multirow{1}{*}{$\sp(2)$}\\\hline
ii.&\includegraphics{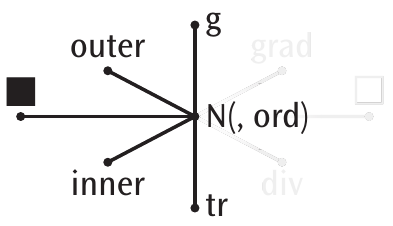} & Any background without $\ord$ ($\xivec$ is then an arbitrary vector field); hyperhomothety with $\ord$ & Subalgebra of $\sp(4)$; maximal parabolic with $\ord$\\\hline
iii.&\includegraphics{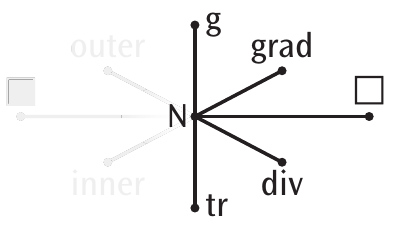} & Symmetric space & Subalgebra of $\sp(4)$; obstruction given by \eqref{eqsp4graddiv}\\\hline
iv.&\includegraphics{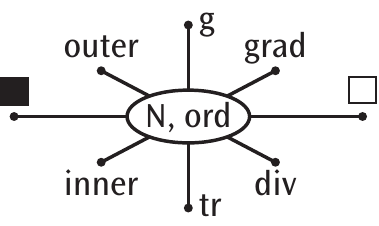} & Hyperhomothety & $\sp(4)$; obstructions given by \eqref{eqsp4graddiv} and \eqref{eqsp4boxgrad}\\\hline
\end{tabular}\
\renewcommand{\arraystretch}{1}
\caption{Root lattices for potential subalgebras of $\sp(4)$, individually conditional upon  hyperhomothety \eqref{eqhomothety} or locally symmetric space \eqref{eqsymmspace} conditions.}\label{figsp4subalgs}
\end{figure}

Thus far, we have simply introduced an operator notation for standard geometric operations on symmetric tensors. Remarkably, subject to combinations of the conditions stipulated in Section~\ref{sectgeometry}, these operators constitute representations of certain deformations of the symplectic Lie algebra $\sp(4)$. 
To see this we first
note that the three operators formed from bilinears in $dx, dx^*$ in Eq.~\eqref{eqsp4ngtr} form a representation of $\sp(2)$ in any background (Figure~\ref{figsp4subalgs}.i) with the index-counting operator $\N$ as its  Cartan generator 
\begin{equation}
  \COM{\N}{\g}=2\,\g,\quad\COM{\N}{\tr}=-2\,\tr,\quad\COM{\g}{\tr}=-4\, (\N+d/2)\, .
\end{equation}
Keeping the  background (and therefore also the vector field $\xivec$) arbitrary, the operators $\N$, $\g$, $\tr$, $\inner$, $\outer$, and $\abox$ form a subalgebra of $\sp(4)$. To add $\ord$ to the algebra as the other $\sp(4)$ Cartan generator, we need to invoke hyperhomothety, and then arrive at a maximal parabolic subalgebra (Figure~\ref{figsp4subalgs}.ii).

Next, relaxing the hyperhomothety condition and instead imposing the locally symmetric space condition~\eqref{eqsymmspace}, we have $\N$, $\g$, $\tr$, $\grad$, $\div$, and $\boxop$ forming a representation of a subalgebra of $\sp(4)$ (Figure~\ref{figsp4subalgs}.iii), up to a mild curvature obstruction given by
\begin{equation}\label{eqsp4graddiv}
  \COM{\div}{\grad}=\boxop-\frac{1}{2}R_{\mu\nu\rho\sigma}M^{\mu\nu}M^{\rho\sigma}\;\,\neq\boxop.
\end{equation}
This computation was performed already in~\cite{Hallowell:2007qk}. Importantly, note
that the operators $\N$, $\g$, $\tr$, $\grad$, $\div$ are all symmetries, in the the sense that
they commute with the operator $\boxop$ which can be interpreted as the Hamiltonian.
In the case of constant curvature manifolds, the obstruction term $\frac{1}{2}R_{\mu\nu\rho\sigma}M^{\mu\nu}M^{\rho\sigma}$ to ($\N$, $\g$, $\tr$, $\grad$, $\div$, $\boxop$) forming a Lie algebra
equals the Casimir operator of the $\sp(2)$ subalgebra built from  ($\N$, $\g$, $\tr$).
In this case it is possible to reformulate this deformed Lie algebra as a novel associative Fourier--Jacobi
 algebra (see~\cite{Hallowell:2007zb}).

Our ultimate algebra requires hyperhomothety, and with all ten operators from Eqs.~(\ref{eqsp4ngtr}--\ref{eqsp4conformalops}) forms a  representation of $\sp(4)$ (Figure~\ref{figsp4subalgs}.iv), up to obstructions given by Eq.~\eqref{eqsp4graddiv} together with
\begin{equation}\label{eqsp4boxgrad}
  \COM{\boxop}{\grad}=-\frac{1}{4}\left(\nabla_\lambda R_{\mu\nu\rho\sigma}\right) \left( M^{\mu\nu}dx^\lambda M^{\rho\sigma}\right)\;\,\neq 0,
\end{equation}
as well as a similar result for $\boxop$ with $\div$. This result is new and its generalization to
more general tensors and spinors is given in Section~\ref{sectosp}.
It is actually valid in any (pseudo)Riemannian background (regardless of whether it is hyperhomothetic
or not). Unlike the obstruction in the Weitzenbock-type identity~\eqref{eqsp4graddiv},
it restricts the symmetries of $\boxop$---regarded as a Hamiltonian---to ($\N$, $\g$, $\tr$)
(and also $(\ord,\inner,\outer,\abox)$ if one considers time-dependent conformal symmetries).
 Note that the right hand side of Eq.~\eqref{eqsp4boxgrad}  vanishes in symmetric spaces, and that both curvature obstructions  vanish together exclusively \emph{in flat space}, where $\sp(4)$ is fully realized as a Lie algebra; any relaxation of flat space to more general backgrounds immediately yields a deformation of $\sp(4)$.
The $\sp(4)$ root lattice and our identification of roots with differential geometry
operators are displayed in Figure~\ref{figsp4rootlatt}. The various different subalgebras discussed above are exhibited  in Figure~\ref{figsp4subalgs}. All possible commutation relations of these operators are tabulated in Appendix~\ref{appcomms}. 

\subsection{Quantum Mechanics}\label{sectsp4qm}
Next we would like to interpret the geometric system described in Section~\ref{sectsp4opsymmtens} as a quantum mechanical one, with the symmetric tensors becoming wavefunctions of particle model with internal structure:
\begin{equation}
  \Phi(x^\mu, dx^\mu)\rightarrow |\Psi\rangle.
\end{equation}
We moreover interpret the coordinate differentials $dx$ and their duals $\dbddx{}$ from the preceding section as raising and lowering operators, respectively:
\begin{displaymath}
  dx^{\mu}\rightarrow a^{\dagger\mu},\quad \dbddx{\mu}\rightarrow a_\mu.
\end{displaymath}
For example, a vector field $\psi_\nu$ can be written in this way as a  wavefunction 
\begin{equation}
  |\Psi\rangle=\psi_\nu(x)a^{\dagger\nu}|0\rangle,
\end{equation}
where we have introduced the Fock vacuum state $|0\rangle$ annihilated by $a_\mu$.

The Heisenberg algebra~\eqref{heis} now 
becomes the standard oscillator one for $a^\dagger$ and $a$ 
\begin{equation}
  \COM{\vphantom{p_\mu x^\mu}a^{\mu}}{a^\dagger_\nu}=\delta^\mu_\nu,
\end{equation}
and similarly the momentum $p_\mu=-i\partial_\mu$ and position $x^\mu$ obey:
\begin{equation}
  \COM{p_\mu}{x^\nu}=-i\delta^\nu_\mu.
\end{equation}
Also, just as the partial derivative is replaced by the canonical momentum, so too the operator $D_\mu$ from Eq.~\eqref{eqsp4D} is replaced by the covariant canonical momentum
\begin{equation}
  -iD_\mu\rightarrow \pi_\mu=p_\mu+i\Gamma_{\mu\nu}^\sigma a^{\dagger\nu} a_\sigma.
\end{equation}

The inner product for our Hilbert space is implied by the normalization  $\langle0|0\rangle=1$, so for example the norm of a pair of eigenstates of $\N$ is
\begin{equation}
  \langle\Phi\,|\Psi\rangle=n!\ \int{\phi}^{*\rho_1\rho_2\ldots\rho_n}\psi_{\rho_1\rho_2\ldots\rho_n}\sqrt{-g}\ d^dx\, .
\end{equation}

The symmetric-tensor operators from the previous section are now manifested as quantized Noether charge operators acting on quantum states, with the appropriate replacements for $dx$, $\dbddx{}$, and $D$ given above. Note that all orderings of operators are fixed by their geometric ancestors.
The algebra satisfied by each charge in this system is exactly identical to that of its geometric \emph{vis-\`a-vis} from Section~\ref{sectsp4opsymmtens}. To gain further insight, we next analyze the classical system underlying this quantum mechanical model.

\subsection{Classical Mechanics}\label{sectsp4actionsymm}
We now work in the classical theory for a particle, where we now choose to interpret the coordinate differentials $dx$ and their duals $\dbddx{}$ from the geometrical representation in Section~\ref{sectsp4opsymmtens}, or alternatively the quantum mechanical raising and lowering operators $a^\dagger$, $a$ from Section~\ref{sectsp4qm}, as comprising a complex-valued vector carried by the particle, i.e.,
\begin{equation}
  a^{\dagger\mu}\;\rightarrow\;\bar{z}^\mu,\quad\; a_\mu\;\rightarrow \;z_\mu.
\end{equation}

We now work in classical theory, so that the quantum mechanical commutators become instead Poisson brackets:
\begin{equation}\label{eqPB}
  \PB{p_\mu}{x^\nu}=\delta_\mu^\nu,\quad\PB{\vphantom{p_\mu x^\mu}\bar{z}^\mu}{z_\nu}=-i\delta^\mu_\nu.
\end{equation}
Throughout this section we assume the hyperhomothety condition \eqref{eqhomothety}. At this point it is convenient to define the covariant variation $\covarvar$ and covariant worldline derivative $\nabla/dt$ as\footnote{
We also note two important identities:
\begin{align*}
\covarvar\dot{x}^{\mu}&=\covartime{}\delta x^{\mu}\\
\COM{\covarvar}{\covartime{}}v^{\mu} &= \delta x^{\rho}\dot{x}^{\sigma}R_{\rho\sigma}{}^{\mu}{}_{\nu}v^{\nu}.
\end{align*}}
\begin{alignat}{2} \label{eqsp4covarvar}
\covarvar v^{\mu}          & =  \delta x^{\sigma}\nabla_{\sigma}v^{\mu}   &&=  \delta v^{\mu} + \Gamma^{\mu}_{\nu\rho} \delta x^{\nu}v^{\rho}\\
\covartime{v^\mu}    & =  \dot{x}^{\sigma}\nabla_{\sigma}v^{\mu}   &&=  \dot{v}^{\mu} + \Gamma^{\mu}_{\nu\rho} v^{\nu}\dot{x}^{\rho}.
\end{alignat}
Since it is central (up to obstructions), we take as our Hamiltonian $H=-\boxop/2$, dropping the $g^{\mu\nu}\Gamma_{\mu\nu}^\sigma D_\sigma$ quantum ordering term for the classical system:
\begin{equation}
  H=\frac{1}{2}\left(\pi^\mu\pi_\mu-R_{\mu}{}^\nu{}_\rho{}^\sigma\bar{z}^\mu z_\nu \bar{z}^\rho z_\sigma\right).
\end{equation}

From \eqref{eqPB} it is evident that we already have Darboux coordinates $x$ and $p$, $z$ and $\bar{z}$, so we can perform a Legendre transformation to obtain a suitable action principle for our particle:
\begin{equation}\label{eqsp4action}
S=\int\left(\frac{1}{2}g_{\mu\nu}\dot{x}^{\mu}\dot{x}^{\nu}+i\bar{z}^{\mu}\frac{\nabla z_{\mu}}{dt} +\frac{1}{2}R_{\,\mu}{}^{\,\nu}{}_{\,\rho}{}^{\,\sigma}\,\bar{z}^{\mu}\,z_{\nu}\,\bar{z}^{\rho}\,z_{\sigma}\right)dt.
\end{equation}
The three terms in this Lagrangian all have interesting geometric interpretations. The first is the usual energy integral, the extremization of which yields simple parametrized geodesic motion. The second ensures parallel transport of the vector $z_\mu$. The third is in effect a coupling between the first two; including it in our model results in many more symmetries (discussed below)  than if it is omitted.

The ten operators forming our representation of $\sp(4)$ in Section~\ref{sectsp4opsymmtens}, listed in Table~\ref{tablesp4opsymmtens}, correspond via Noether's theorem to conserved quantities of the above action, with explicit symmetries listed in Table~\ref{tablesp4actionsymm}. There are, however, three interesting caveats to make. First, because there is an obstruction to the algebra between the $\boxop$ charge and both the $\grad$ and $\div$ charges (Eq.~\eqref{eqsp4boxgrad}), $\grad$ and $\div$ technically \emph{are not} conserved charges of our action---performing the $\grad$ variation, we find
\begin{equation}
  \delta S=i\!\int\left(\nabla_\lambda R_{\mu}{}^\nu{}_\rho{}^\sigma\bar{z}^\lambda\bar{z}^\mu z_\nu \bar{z}^\rho z_\sigma\right)dt
  \quad\text{subject to}\quad\delta x^{\mu}=i\bar{z}^{\mu},\;\covarvar z_{\mu}=\dot{x}_{\mu}.
\end{equation}
The result of this variation, which is exactly the classical equivalent of Eq.~\eqref{eqsp4boxgrad}, shows the failure of the $\grad$ and $\div$ symmetries due to the curvature of the underlying manifold.

Second, it is important to note that the symmetries corresponding to the hyperhomothetic operators---$\ord$, $\inner$, $\outer$, and $\abox$---do not commute with the Hamiltonian~$H$; consequently, they are off-shell symmetries and furthermore possess \emph{time dependence;} it is of course sufficient to set $t=0$, since they are conserved. However, because of the subalgebra of the $\sp(4)$ Lie algebra which they satisfy, their time dependence can be computed analytically, through the following procedure. First, given a quantized Noether charge operator $Q_0$, its time dependence is determined using the system's Hamiltonian operator $H$:
\begin{equation}
  \frac{dQ}{dt}=\COM{Q_0}{H},
\end{equation}
which can be integrated to yield $Q(t)$. For example, to obtain the time dependence of the $\ord$ charge\begin{equation}
  \frac{d}{dt}\ord=\COM{\ord}{H}=2H\quad\Rightarrow\quad\ord(t)=\ord+2H t,
\end{equation}
or for $\abox$,
\begin{equation}
  \frac{d}{dt}\abox=\COM{\abox}{H}=-4\ord\quad\Rightarrow\quad\abox(t)=\abox-4\ord\, t-4\boxop t^2.
\end{equation}
Notice that the spectrum-generating algebra obeyed by the charges allows
us to explicitly integrate their time dependence. From here, $Q$'s corresponding rigid symmetry for the coordinate $q^{\mu}$ is then
\begin{equation}
\delta_{Q}q^{\mu}=\COM{Q(t)}{q^{\mu}}.
\end{equation}
Proceeding in this way one can obtain all the symmetries from Table~\ref{tablesp4actionsymm}.


\begin{table}[htbp]
\renewcommand{\arraystretch}{1.5}
\newcommand{\commonphantom}{\ \phantom{\covarvar\bar{z}^{\mu}}}
\newcommand{\specalign}[1]{\commonphantom\mathllap{#1}}
\centering \begin{tabular}{|c|l|c|}
\hline \it Name & \multicolumn{1}{c|}{\it Symmetry} & \it Noether charge\\\hline
\multirow{2}{*}{$\N$} & $\commonphantom\mathllap{\delta z_{\mu}}=-z_{\mu}$ & \multirow{2}{*}{$\bar{z}^{\mu}z_{\mu}$}\\
& $\specalign{\delta \bar{z}^{\mu}}=\bar{z}^{\mu}$ & \\\hline
$\g$        & $\specalign{\delta z_{\mu}}=g_{\mu\nu}\bar{z}^{\nu}$ & $g_{\mu\nu}\bar{z}^{\mu}\bar{z}^{\nu}$\\\hline
$\tr$       & $\specalign{\delta \bar{z}^{\mu}}=g^{\mu\nu}z_{\nu}$ & $g^{\mu\nu}z_{\mu}z_{\nu}$\\\hline
\multirow{2}{*}{$\grad$}     & $\specalign{\delta x^{\mu}}=i\bar{z}^{\mu}$ & \multirow{2}{*}{$\dot{x}^{\mu}\bar{z}_{\mu}$}\\
 & $\specalign{\covarvar z_{\mu}}=\dot{x}_{\mu}$ & \\\hline
\multirow{2}{*}{$\div$} & $\specalign{\delta x^{\mu}}=i\bar{z}^{\mu}$ & \multirow{2}{*}{$\dot{x}^{\mu}z_{\mu}$}\\
& $\specalign{\covarvar \bar{z}^{\mu}}=\dot{x}^{\mu}$ &\\\hline
\multirow{3}{*}{$\boxop$}     & $\specalign{\delta x^{\mu}}=\dot{x}^{\mu}$ & \multirow{3}{*}{$\frac{1}{2}\dot{x}^{\mu}\dot{x}_{\mu} - \tfrac{1}{2}R_{\mu}{}^{\nu}{}_{\rho}{}^{\sigma}\bar{z}^{\mu}z_{\nu}\bar{z}^{\rho}z_{\sigma}$}\\
 & $\specalign{\covarvar \bar{z}^{\mu}}=-iR_\alpha{}^\mu{}_\gamma{}^\delta\,\bar{z}^\alpha\,\bar{z}^\gamma\, z_\delta$ & \\
 & $\specalign{\covarvar z_{\mu}} = iR_\mu{}^\beta{}_\gamma{}^\delta\, z_\beta\,\bar{z}^\gamma z_\delta$ & \\\hline
$\ord$      &  $\specalign{\delta x^{\mu}}=2t\dot{x}^{\mu} - \xi_{\mu}$  & $\dot{x}^{\mu}\xi_{\nu} - t\dot{x}^{\mu}\dot{x}_{\mu}$\\\hline
\multirow{2}{*}{$\inner$}     & $\specalign{\delta x^{\mu}}=-itz^{\mu}$ & \multirow{2}{*}{$x^{\mu}z_{\mu}-t\dot{x}^{\mu}z_{\mu}$}\\
 & $\specalign{\covarvar \bar{z}^{\mu}}=t\dot{x}^{\mu}-\xi^{\mu}$ & \\\hline
\multirow{2}{*}{$\outer$}     & $\specalign{\delta x^{\mu}}=i t \bar{z}^{\mu}$ & \multirow{2}{*}{$x^{\mu}\bar{z}_{\mu}-t\dot{x}^{\mu}\bar{z}_{\mu}$}\\
 & $\specalign{\covarvar z_{\mu}}=t\dot{x}_{\mu}-\xi_{\mu}$ & \\\hline
$\abox$     &  $\specalign{\delta x^{\mu}} = t^2\dot{x}^{\mu} + t \xi^{\mu}$  & $\xi^{\mu}\xi_{\mu} - 2t\dot{x}^{\mu}\xi_{\mu} + t^2\dot{x}^{\mu}\dot{x}_{\mu}$\\\hline
\end{tabular}
\renewcommand{\arraystretch}{1}
\caption{Symmetries and corresponding Noether charges of the action from \eqref{eqsp4action}. Unspecified variations are zero.}\label{tablesp4actionsymm}
\end{table}

\section{$\osp(Q|2p+2)$ Conformal Quantum Mechanics}\label{sectosp}
We now generalize the algebra from Section~\ref{sectsp4opsymmtens} to the  orthosymplectic Lie superalgebra $\osp(Q|2p+2)$. We will always assume the hyperhomothety condition from Eq.~\eqref{eqhomothety} in this section and use  the spin connection $\omega_{\mu}{}^m{}_n$ rather than the Christoffel symbols to build covariant derivative operators. In place of the coordinate differentials $dx^{\mu}$ and $\dbddx{\mu}$ from before, we now employ an orthosymplectic vector denoted by $X_{\alpha}$, where~$\alpha$ is a superindex taking values $1\leq\alpha\leq 2p+Q$. For $1\leq\alpha\leq 2p$, $X_{\alpha}$ is a bosonic variable, and otherwise it is fermionic. These  satisfy the  supercommutation relations:
\begin{equation}
  \SCOM{X^{m}_{\alpha}}{X^{n}_{\beta}}=\eta^{mn}J_{\alpha\beta},
\end{equation}
Here $J_{\alpha\beta}$ is the orthosymplectic bilinear form, given by
\begin{equation}
    \left(J_{\alpha\beta}\right)=\begin{cases}\vspace*{.2cm}\left(\begin{array}{cc|cc}&-\mathbb{1}_{p\times p}&&\\\mathbb{1}_{p\times p}&&&\\\hline&&&\mathbb{1}_{q \times q}\\&&\mathbb{1}_{q\times q}&\end{array}\right)& \text{$Q=2q$ even,}\\
    \left(\begin{array}{cc|ccc}&-\mathbb{1}_{p\times p}&&&\\\mathbb{1}_{p\times p}&&&&\\\hline&&&\mathbb{1}_{q \times q}&\\&&\mathbb{1}_{q\times q}&&\\&&&&1\end{array}\right)& \text{$Q=2q+1$ odd.}
  \end{cases}
\end{equation}
We denote the inverse of $J_{\alpha\beta}$ by $J^{\beta\alpha}$, so that $J_{\alpha\beta}J^{\gamma\beta}=\delta^\gamma_\alpha$ (while $J^{\beta\alpha}J_{\gamma\beta}=-T^{\alpha\kappa}_{\kappa\gamma}$; see Appendix \ref{appT}).

From the $X$'s, we can now form the $SO(d)$ rotation/Lorentz generators
\begin{equation}\label{rotation}
  M^{mn}=J^{\beta\alpha}X_{\vphantom{\beta}\alpha}^{[m} X_{\beta}^{n]},
\end{equation}
as well as the covariant derivative operator:
\begin{equation}\label{eqospD}
  D_\mu=\partial_\mu+\omega_{\mu mn}M^{mn}.
\end{equation}

Next, from the space of orthosymplectic bilinears, we define
\begin{equation}
  f_{\alpha\beta}=\eta_{mn}X_{(\alpha}^m X_{\beta]}^n,
\end{equation}
where $(\alpha \beta]$ denotes antisymmetrization if $\alpha$ and $\beta$ are both fermionic indices, and otherwise denotes symmetrization; we again refer the reader to Appendix~\ref{appT}, where a computational scheme for handling such symmetrization is detailed. The supermatrix $f_{\alpha\beta}$ subsumes $\N$, $\g$, and $\tr$ from Section~\ref{sectsp4opsymmtens}, as detailed below.

In turn we introduce dynamical symmetry generators
\begin{equation}
  v_{\alpha}=X_{\alpha}^m e^{\mu}{}_{m}D_\mu\equiv X_{\alpha}^{m}D_{m},
\end{equation}
which subsumes $\grad$ and $\div$, and which could be viewed as a generalized Dirac operator. We also clearly need
a Lichnerowicz wave operator 
\begin{equation}
  \boxop = D^m D_m-\omega^{nm}{}_nD_m+\frac{1}{4}R_{mnrs}M^{mn}M^{rs}.
\end{equation}

To complete our set, we now add the hyperhomothetic operators. The operators~$\ord$ and~$\abox$ remain unchanged, except of course that $\ord$ now uses the orthosymplectic covariant derivative operator from Eq.~\eqref{eqospD}:
\begin{equation}
  \ord = -\xi^m D_m,\quad  \abox = \xi^m \xi_m.
\end{equation}
Lastly,  the generalizations of $\inner$ and~$\outer$ are:
\begin{equation}
  w_{\alpha}=\xi_m X_\alpha^m.
\end{equation}

The explicit correspondence between the $\osp(Q|2p+2)$ operators ($f_{\alpha\beta}$, $v_\alpha$, $w_\alpha$) and their $\sp(4)$ equivalents is
\begin{gather}
  \left(f_{\alpha\beta}\right)\leftrightarrow\left(\begin{array}{cc}\g&\!\!\!\N+d/2\! \!\!\\ \!\!\! \N+d/2\!\!&\tr\end{array}\right),\:\:  \left(v_{\alpha}\right)\leftrightarrow\left(\begin{array}{cc}\grad\\\div\end{array}\right),\;\: \left(w_{\alpha}\right)\leftrightarrow\left(\begin{array}{cc}\outer\\\inner\end{array}\right).
\end{gather}

The symmetry algebras which can be formed by our $\osp(Q|2p+2)$ operators are exactly analogous to the $\sp(4)$ case (Figure~\ref{figsp4subalgs}), with $\sp(2)\rightarrow\osp(Q|2p)$ and $\sp(4)\rightarrow\osp(Q|2p+2)$. Specifically:
\begin{enumerate}
  \item $f_{\alpha\beta}$ form $\osp(Q|2p)$ in any background;
  \item $f_{\alpha\beta}$, $w_\alpha$, and $\abox$ form a subalgebra of $\osp(Q|2p+2)$ in any background, which becomes maximal and parabolic with the addition of $\ord$ and the hyperhomothety condition (Eq.~\eqref{eqhomothety});
  \item $f_{\alpha\beta}$, $v_\alpha$, and $\boxop$ form a subalgebra of $\osp(Q|2p+2)$ under the locally symmetric space condition, Eq.~\eqref{eqsymmspace}, with an obstruction~\eqref{eqospvv};
  \item together $f_{\alpha\beta}$, $\ord$, $v_\alpha$, $\boxop$, $w_\alpha$, and $\abox$  form $\osp(Q|2p+2)$, subject to obstructions \eqref{eqospboxv} and \eqref{eqospvv}.
\end{enumerate}
We now present the two obstructions to the full $\osp(Q|2p+2)$ algebra under hyperhomothety, respectively generalizing their $\sp(4)$ counterparts in Eqs.~\eqref{eqsp4graddiv} and \eqref{eqsp4boxgrad}:
\begin{alignat}{2}
  \label{eqospvv}\SCOM{v_\alpha}{v_\gamma}&=J_{\alpha\gamma}\left(\omega_m{}^{mr}D_r+D^rD_r\right)+\frac{1}{2}R_{mnrs}X_\alpha^m X_\gamma^n M^{rs}&\;\,&\neq J_{\alpha\gamma}\boxop,\\
  \label{eqospboxv}\COM{\boxop}{v_{\alpha}}&=-\frac{1}{4}\left(\nabla_t R_{mnrs}\right)\left(M^{mn}X_{\alpha}^tM^{rs}\right)&\;\,&\neq 0.
\end{alignat}
Both these obstructions can be removed when $Q=0,1$ and $p=0$ at which values our models revert to ${\cal N}=1,2$
supersymmetric quantum  mechanics (see~\cite{Hallowell:2007qk} for details).
The remainder of the algebra is explicitly presented in Appendix~\ref{appcomms}.

The Hilbert space of these othosymplectic models is described in great detail in~\cite{Hallowell:2007qk} (various special cases have been studied in~\cite{Duval,DuboisViolette:1999rd,Bekaert:2003az,Olver,edgar,de Medeiros:2003dc,Bastianelli:2007pv}). Wavefunctions are tensors  
expanded in terms of multi-forms and multi-symmetric-forms. Moreover, when $Q$ is odd, wavefunctions are spinor-valued so carry a spinor index $\alpha$.
In a Young diagram notation, where rows are totally symmetric and columns antisymmetric,  we would write
\be
\Phi^\alpha{}_{\tiny\Yvcentermath1 
\!\!\!\!\!\!\!\!\!\!\!\!\!\!\!\!\!\!\mbox{$p$ times}
\left\{\!\!\!
\begin{array}{c}\yng(4)\\ \otimes \\ \yng(5)\\ \otimes \\[-1mm] \vdots \\ \otimes \\ \yng(3)\end{array}
\right.
\otimes \ \
\underbrace{\yng(1,1,1,1)\otimes\yng(1,1,1)\otimes\cdots\otimes\yng(1,1,1,1,1)}_{\mbox{$[Q/2]$ times}}}\, .
\label{tensors}
\ee
Clearly, although this is not an irreducible basis for tensors and spinors on a manifold, we can
generate all such objects this way. Irreducible tensors
can be obtained by placing constraints coming from parabolic subgroups of the $\osp(Q|2p)$ algebraically
acting $R$-symmetry algebra generated by $f_{\alpha\beta}$.

\section{Three Dimensions}
\label{three}
All our results so far have not depended crucially on the 
dimen\-sionality of the background manifold. In this section 
we wish to study particle symmetries built from the totally
antisymmetric Levi-Civita symbol which are 
dimen\-sion-depen\-dent. Therefore, in this initial study, we 
concentrate on three dimensions and the Noether charges that
can be built from the three-dimensional Levi-Civita symbol\footnote{In our notations the Levi-Civita symbol is a density so $\varepsilon^{\mu\nu\rho}$ is characterized by $\varepsilon^{123}=1$ in any coordinate system. } $\varepsilon^{\mu\nu\rho}$.
We further specialize to constant curvature backgrounds with the advantage of maximizing
the set of symmetries.

To be completely concrete let us take the three dimensional hyperbolic  metric
\begin{gather}\label{ds2}
ds^2 = \frac{dx^2+dy^2+dz^2}{z^2}\, ,
\end{gather}
although none of our algebraic results depend on this choice of coordinates.
We focus on the $sp(2)$ symmetric tensor model, but the generalization to $osp(Q|2p)$
spinning degrees of freedom is immediate. Representing $dx_\mu^*=\partial/\partial(dx^\mu)$
acting on symmetic tensors viewed as analytic functions of commuting differentials $dx^\mu$, the
covariant derivative operator $D_\mu$ of~\eqref{eqsp4D} is given explicitly by
\begin{alignat}{2}
D_x=&\
\partial_x
-\frac1z
\Big\{
dx\frac{\partial}{\partial(dz)}-dz\frac{\partial}{\partial(dx)}
\Big\}
\, , \nonumber\\
D_y=&\
\partial_y
-\frac1z
\Big\{
dy\frac{\partial}{\partial(dz)}-dz\frac{\partial}{\partial(dy)}
\Big\}
\, ,\nonumber\\
D_z=&\
\partial_z+\frac1z \, \N\, ,
\end{alignat}
where the index-counting operator 
\begin{equation}
\N=dx\frac{\partial}{\partial(dx)}+dy\frac{\partial}{\partial(dy)}+dz\frac{\partial}{\partial(dz)}\, .
\end{equation}
The $sp(2)$ algebra of internal symmetries is generated by $\N$, $\g=ds^2$ as in~\eqref{ds2}, and
the trace operator
\begin{equation}
\tr=z^2\Big\{\frac{\partial^2}{\partial(dx)^2}
+\frac{\partial^2}{\partial(dy)^2}
+\frac{\partial^2}{\partial(dz)^2}
\Big\}\, .
\end{equation}
The quadratic $sp(2)$ Casimir is
\be
\cas=\g\,  \tr -\N(\N+1)\, .
\ee
Then we have the pair of differential operators
\begin{alignat}{2}
\grad=& \ dx D_x+dy D_y+dz D_z \, ,\nonumber\\
\div=& \ z^2\Big\{\frac{\partial}{\partial(dx)} D_x
+\frac{\partial}{\partial(dy)} D_y+\frac{\partial}{\partial(dz)} D_z\Big\}\, ,
\end{alignat}
which form an $sp(2)$ doublet. In turn, the Lichnerowicz wave operator
\be
\Box=\div \ \grad-\grad\ \div -2\,\cas\, 
\ee
is central. So far we have simply written out the results of the previous sections
explicitly for this three dimensional background. Now we introduce a new operator 
built from the Levi-Civita symbol
\be
\curl = \frac{1}{\sqrt{g}}\, \varepsilon^{\mu\nu\rho} dx_\mu \frac{\partial}{\partial(dx^\nu)}D_\rho
=z\Big\{
dx \frac{\partial}{\partial(dy)}D_z
\pm \mbox{ permutations}
\Big\}\, .
\ee
Geometrically, this is the symmetrized curl operation first introduced in~\cite{Damour:1987vm}. It can be generalized 
to more spinning degrees of freedom by considering
\be
\curl = \frac{1}{2\sqrt{g}}\, \varepsilon^{\mu\nu\rho} M_{\mu\nu}D_\rho\, ,
\ee
where the $SO(3)$ rotation operators are given in~\eqref{rotation}.
The symmetrized curl operator has various interesting properties.
Firstly, it is central, i.e. it commutes with $\{\Box,\div,\grad,\tr,\N,\g\}$.
Moreover, since it is central, so is its square which equals
\be
\curl^2=\Box(\g\,\tr - \N^2) + \grad(2\N+1)\div - \g\,\div^2 - \grad^2\, \tr+\cas(\cas+2\N)\, .
\ee
This operator is rather interesting: It has been known for quite some time that
the Fourier--Jacobi algebra obeyed by $\{\Box,\div,\grad,\tr,\N,\g\}$ in {\it flat} backgrounds possessed a 
quadratic Casimir~\cite{Berndt}. The above result shows that this Casimir can be generalized
to constant curvature backgrounds (at least in three---and possibly higher---dimensions).

Another interesting consequence follows by considering the quantum mechanical origin of the
above operators. In our previous models, we viewed the above operators as Noether charges
and took $\Box$ as the Hamiltonian to ensure a maximal set of symmetries. In three dimensions
however, we have the additional possibililty of regarding the curl operator as the Hamiltonian.
This corresponds to a new particle model with action
\be
S=\int \left(\pi_\mu \dot x^\mu  +i \bar z^\mu\frac{\nabla z_\mu}{dt}
-\frac{1}{\sqrt{g}}\ \varepsilon^{\mu\nu\rho} \bar z_\mu z_\nu \pi_\rho \right)dt \, ,
\ee
and symmetries corresponding to the above Noether charges. Notice that this model does not admit
a Legendre transformation with respect to $\pi_\mu$, but if a second-order theory is desired, one can 
always add an additional term to the Hamiltonian proportional to the Lichnerowicz wave operator.
Finally, we note that a gauged version of this model should provide a first-quantized
description of three topologically massive theories~\cite{Deser:1982vy}.

\section{Conclusions}
In a previous work~\cite{Hallowell:2007qk} 
it was found that Weitzenbock identities for differential operators acting on tensors of general types on symmetric spaces
followed from  a symmetry algebra that formed the maximal parabolic {\it subalgebra} of a larger orthosymplectic algebra. These symmetries
could also be realized as generalized supersymmetries and $R$-symmetries of an underlying curved space quantum mechanical model.
A puzzle remained, however---namely whether the {\it full} orthosymplectic algebra could be realized as symmetries of the quantum mechanical
model and therefore, in turn, as symmetries of differential operators on manifolds. In this paper we have solved this puzzle by showing that
the remainder of the orthosymplectic algebra corresponds to worldline conformal symmetries of the quantum mechanical model. These 
symmetries are unobstructed by curvatures when the background geometry possesses a hypersurface-orthogonal conformal
Killing vector. In terms of geometry, the new conformal symmetries correspond to all operations on tensors that can be performed
using this vector.

Another aspect of the work~\cite{Hallowell:2007qk} was that quantum mechanical symmetries were built from all available invariant tensors save the totally antisymmetric
Levi-Civita symbol. In the original work of~\cite{Damour:1987vm}, in the context of topologically massive theories, it was also realized that a symmetrized curl operation could be added to the algebra of operators acting on symmetric tensors in three dimensions. Clearly this operation can be generalized
to tensors of different types. Furthermore, there should be a quantum mechanical model underlying this symmetry. We have constructed 
exactly this model in the  present paper.

An immediate application of our results is to projective geometry. If we consider a model
which enjoys the full orthosymplectic algebra as its symmetry group, we can take the quadratic Casimir as Hamiltonian. In flat space this amounts taking the square of the total angular momentum operator as Hamiltonian. By examining the square of the orbital angular momentum operator
$M_{\mu\nu}^{2}=x^{2} p^{2} - x\cdot p^{2}$ we see that the Hamiltonian $H\sim x^{2} \widehat p^{2}$ where the covector $\widehat p$ is
the momentum orthogonal to the homothety $x$. This model therefore describes motion on the projective space  \raisebox{-.3mm}{$\mathbb R$}P$^{d-1}$  by projectivizing with respect to dilations. It would be most interesting to generalize this construction to the models considered here and obtain an algebra of operators acting on projective spaces.

\section*{Acknowledgements}
The authors are pleased to thank Karl Hallowell for an important contribution. J.~B. was supported, in part, by VIGRE grant DMS-0636297
and appreciates  its role encouraging undergraduate research.

\appendix
\section{Supercommutation Computation}\label{appT}
A  convenient method for supercommutator and superindex computations is to define a symbol $T^{\mu\nu}_{\alpha\beta}$, similar to a four-index Kronecker delta, which when contracted  over  $\mu$ and $\nu$ produces a sign depending on the values of the indices  $\alpha$ and $\beta$. The sign symbol $T$ allows us to use standard Einstein notation for tensors  and is easily implemented in computer algebra applications. 
Specifically, we define
\begin{equation}
T^{\delta\epsilon}_{\alpha\beta}=\begin{cases}
0& \text{if $\alpha\neq\delta$ or $\beta\neq\epsilon$},\\
-1& \text{if $(\alpha=\beta)>2p$},\\
+1& \text{otherwise}.
\end{cases}
\end{equation}
The Leibniz rule for supercommutators (here we are working in $\osp(Q|2p)$ with generators $X_\alpha$, where indices are orthosymplectic superindices---see Section~\ref{sectosp}) is then simply
\begin{equation}
  \SCOM{X_{\alpha}}{X_{\beta}X_{\gamma}} = \SCOM{X_{\alpha}}{X_{\beta}} X_\gamma + T^{\delta\epsilon}_{\alpha\beta} X_\epsilon\SCOM{X_\delta}{X_\gamma}\, ,
\end{equation}
while for supercommutators of bilinears we find
\begin{equation}
  \SCOM{A_{\alpha\beta}}{X_{\gamma}X_\delta}=\SCOM{A_{\alpha\beta}}{X_\gamma}X_\delta+T^{\mu\rho}_{\alpha\sigma}T^{\nu\sigma}_{\beta\gamma} X_\rho\SCOM{A_{\mu\nu}}{X_\delta}.
\end{equation}
Symmetry or antisymmetry in two orthosymplectic indices depending on their value reads
\begin{equation}
  A_{(\alpha}B_{\beta]}=\frac{1}{2}\left(A_\alpha B_\beta+T^{\delta\epsilon}_{\alpha\beta}A_\epsilon B_\delta\right)
\end{equation}
($T$ replaces the standard $(-)_{\alpha\beta}$ symbol).
The following set of identities are most useful:
\begin{align*}
  T^{\mu\nu}_{\alpha\gamma}T^{\rho\sigma}_{\mu\nu}&=\delta^\rho_\alpha\delta^\sigma_\gamma,&
  J_{\alpha\beta}T^{\beta\kappa}_{\kappa\gamma}&=-J_{\gamma\alpha},&
  T^{\mu\nu}_{\beta\gamma}J_{\alpha\nu}&=T^{\mu\nu}_{\beta\alpha}J_{\nu\gamma},\\
  \tau^{\mu\nu}_{\alpha\beta}J_{\mu\nu}&=-J_{\beta\alpha},&
  J^{\beta\alpha}J_{\gamma\beta}&=-T^{\alpha\kappa}_{\kappa\gamma},&
  J^{\beta\alpha}J_{\beta\gamma}&=\delta^{\alpha}_\gamma,\\
  T^{\omega\lambda}_{\gamma\delta}T^{\nu\sigma}_{\beta\lambda}T^{\mu\rho}_{\alpha\sigma}J_{\nu\omega}&=T^{\mu\rho}_{\alpha\delta}J_{\beta\gamma},&
  T^{\epsilon\kappa}_{\beta\gamma}T^{\beta\zeta}_{\zeta\kappa}&=\delta^\epsilon_\gamma,&
  T^{\mu\nu}_{\alpha\beta}T^{\sigma\rho}_{\nu\gamma}J_{\mu\rho}&=\delta^\sigma_\beta J_{\alpha\gamma}.
\end{align*}

\section{Explicit Algebras}\label{appcomms}
\newcommand{\altsign}[1]{\underline{#1}}
We present the explicit commutation algebras of the $\sp(4)$ operators from Sections~\ref{sectsp4opsymmtens} and the $\osp(Q|2p+2)$ operators of Section~\ref{sectosp} in Figures~\ref{figappcommssp4} and \ref{figappcommsosp}, respectively. Note that the operators $\altsign{\N}$ and $\altsign{\ord}$ used here are defined by
$
  \altsign{\N}=\N+d/2,\quad\altsign{\ord}=\ord-d/2$, 
where $d$ represents the dimension of the manifold in question. These operators are Lie algebra elements.

\begin{figure}[h]
\newcolumntype{K}{>{\centering\arraybackslash} c }
\renewcommand{\arraystretch}{1.3}

\centering

\begin{tabular}{|K|K|K|K|K|K|K|K|K|K|K|}\cline{2-10}
\multicolumn{1}{c|}{$\COM{\,\cdot\:}{\cdot\,}$}& $\!\ord\!$ & $\g$ & $\tr$ & $\grad$ & $\div$ & $\boxop$ & $\outer$ & $\inner$ & $\abox$\\\hline
$\N$ & 0&$2\g$&$-2\tr$&$\grad$&$-\div$&0&$\outer$&$-\inner$& 0\\\hline
$\ord$  &&0&0&$\!\grad^\dagger\!$&$\div^\dagger$&$2\boxop^\dagger$&$-\outer^\dagger$&$-\inner^\dagger$&$-2\abox^\dagger$\\\hline
$\g$ & &&$-4\altsign{\N}$&0&$\!-2\grad\!$&0&0&$-2\outer$&0\\\hline
$\tr$ &&&&$2\div$&0&0&$2\inner$&0&0\\\hline
$\!\grad\!$  &&&&&\!Eq.~\ref{eqsp4graddiv}\!&Eq.~\ref{eqsp4boxgrad}&$\g^\dagger$&$\!\N+\ord^\dagger\!$&$\outer^\dagger$\\\hline
$\div$  &&&&&&\!Eq.~\ref{eqsp4boxgrad}\!&$\!\altsign{\N}-\altsign{\ord}^\dagger\!$&$\tr^\dagger$&$2\inner^\dagger$\\\hline
$\boxop$  &&&&&&&$2\grad^\dagger$&$2\div^\dagger$&$\!-4\altsign{\ord}^\dagger\!$\\\hline
$\outer$  &&&&&&&&$-\abox$&0\\\hline
$\inner$  &&&&&&&&&0\\\hline
\end{tabular}
\renewcommand{\arraystretch}{1}
\caption{Explicit commutation algebra of the operators from Section~\ref{sectsp4opsymmtens}. All results apply to arbitrary curved manifolds, except those labeled ${}^\dagger$, which are contingent upon application of the hyperhomothety condition from Eq.~\eqref{eqhomothety}. Commutators are of the form $\COM{\text{left column}}{\text{top row}}$. }\label{figappcommssp4}
\end{figure}

\begin{figure}[h]
\newcolumntype{K}{>{\centering\arraybackslash} c }
\renewcommand{\arraystretch}{1.3}
\centering
\begin{tabular}{|K|K|K|K|K|K|K|}\cline{2-7}
\multicolumn{1}{c|}{$\!\SCOM{\,\cdot\:}{\cdot\,}\!$} &$f_{\gamma\delta}$&$\ord$&$v_{\gamma}$&$\boxop$&$w_\gamma$&$\abox$\\\hline
$f_{\alpha\beta}$&$\!4J_{(\beta(\gamma}f_{\alpha]\delta]}\!$&0&$2v_{(\alpha}J_{\beta]\gamma}$&0&$2w_{(\alpha}J_{\beta]\gamma}$&0\\\hline
$\ord$&0&0&$v_\gamma^\dagger$&$2\boxop^\dagger$&$-w_\gamma^\dagger$&$-2\abox$\\\hline
$v_\alpha$&$2J_{\alpha(\gamma}v_{\delta]}$&$-v_\alpha^\dagger$&Eq.~\ref{eqospvv}&Eq.~\ref{eqospboxv}&$\!-J_{\alpha\gamma}\ord+f_{\alpha\gamma}^\dagger\!$&$2w_\alpha^\dagger$\\\hline
$\boxop$&0&$\!-2\boxop^\dagger\!$&Eq.~\ref{eqospboxv}&0&$2v_\gamma^\dagger$&$-4\ord^\dagger$\\\hline
$w_\alpha$&$2J_{\alpha(\gamma}w_{\delta]}$&$w_{\alpha}^\dagger$&$\!-J_{\alpha\gamma}\ord-f_{\alpha\gamma}^\dagger\!$&$-2v_{\alpha}^\dagger$&$J_{\alpha\gamma}\abox$&0\\\hline
\end{tabular}
\renewcommand{\arraystretch}{1}
\caption{Explicit supercommutation algebra of the operators from Section~\ref{sectosp}. All results apply to arbitrary curved manifolds, except those labeled ${}^\dagger$, which are contingent upon application of the hyperhomothety condition from Eq.~\eqref{eqhomothety}. Supercommutators are of the form $\SCOM{\text{left column}}{\text{top row}}$. Note that the symmetrizations on the result for $\SCOM{f_{\alpha\beta}}{f_{\gamma\delta}}$ implies symmetry of the form  $\left(\beta\alpha\right]$ and~$\left(\gamma\delta\right]$.}\label{figappcommsosp}
\end{figure}



\end{document}